\documentclass[fleqn,usenatbib]{mnras}
\usepackage{hyperref}
\usepackage{amsmath}
\usepackage{amssymb,latexsym,mathrsfs}
\usepackage{graphicx}
\usepackage{commath}
\usepackage{dcolumn}
\usepackage{siunitx}
\usepackage{float}
\usepackage{bm}
\usepackage{epsfig}
\usepackage[version=3]{mhchem}
\usepackage{color}
\usepackage{multirow}
\usepackage{makecell}
\usepackage{acro}
\usepackage{textgreek}
\usepackage{caption}
\usepackage[title]{appendix}
\usepackage{tabularx}

\DeclareAcronym{$r_{disk}$}{
  short=$r_{disk}$,
  long= Scale length of galaxy observed at 3.6$\mu m$,
}
\DeclareAcronym{DM}{
  short=DM,
  long=Dispersion Measure,
}
\DeclareAcronym{$DM_{IGM}$}{
  short=$DM_{IGM}$,
  long=Dispersion Measure from Intergalactic Medium,
}
\DeclareAcronym{$DM_{MW}$}{
  short=$DM_{MW}$,
  long=Dispersion Measure from Milky Way,
}
\DeclareAcronym{$DM_{host}$}{
  short=$DM_{host}$,
  long=Dispersion Measure from host galaxy,
}
\DeclareAcronym{$DM_{GalaxyDisk}$}{
  short=$DM_{GalaxyDisk}$,
  long=Dispersion Measure from disk of host galaxy,
}
\DeclareAcronym{$DM_{local}$}{
  short=$DM_{local}$,
  long=Dispersion Measure from local environment of FRB,
}
\DeclareAcronym{$DM_{source}$}{
  short=$DM_{source}$,
  long=Dispersion Measure from intrinsic emission mechanism of FRB,
}
\DeclareAcronym{ms}{
  short=ms,
  long=millisecond,
}
\DeclareAcronym{z}{
  short=z,
  long = Redshift,
}

\DeclareAcronym{FRB}{
  short=FRB,
  long = Fast Radio Burst,
}

\DeclareAcronym{$z_r$}{
  short=$z_r$,
  long = epoch of HeII reionization,
}

\DeclareAcronym{$z_{r, data}$}{
  short=$z_{r, data}$,
  long = epoch of HeII reionization used to generate synthetic FRB data,
}

\DeclareAcronym{$z_{r, fit}$}{
  short=$z_{r, fit}$,
  long = epoch of HeII reionization obtained from the fitting,
}

\hypersetup{
    colorlinks=true,
    linkcolor=red,
    citecolor=blue,
    urlcolor=magenta,
} 

\newcommand{\be}{\begin{equation}}
\newcommand{\ee}{\end{equation}}
\newcommand{\bea}{\begin{eqnarray}}
\newcommand{\eea}{\end{eqnarray}}

\title[Constraining \ce{HeII} Reionization Detection from FRBs]{Constraining \ce{HeII} Reionization Detection Uncertainties via Fast Radio Bursts} 

\author[A.L., A. M \&M.S \& G.S.]{
Albert Wai Kit Lau,$^{1}$
Ayan Mitra,$^{2,3}$\thanks{E-mail: ayan.mitra@nu.edu.kz}
Mehdi Shafiee$^{4,5,6}$
George Smoot$^{1,4,7,8,9}$
\\
$^{1}$Department of Physics, The Hong Kong University of Science and Technology, Clear Water Bay, Kowloon, Hong Kong\\
$^{2}$School of Engineering and Digital Sciences, Nazarbayev University, Nur-Sultan, Kazakhstan\\
$^{3}$Kazakh-British Technical University, Almaty, Kazakhstan\\
$^{4}$Energetic Cosmos Laboratory, Nazarbayev University, Nur-Sultan, Kazakhstan\\
$^{5}$Department of Physics, Engineering Physics Astronomy, Queen’s University, Kingston, ON Canada\\
$^{6}$Arthur B. McDonald Canadian Astroparticle Physics Research Institute, Queen's University, Kingston, ON Canada\\
$^{7}$Paris  Centre  for  Cosmological  Physics,  Universit{\'e}  de  Paris,
CNRS,  Astroparticule  et  Cosmologie,  F-75013  Paris,  France\\
$^{8}$Physics Department and LBNL, University of California, Berkeley, CA, 94720 USA \\
$^{9}$Institute for Advanced Study Hong Kong University of Science and Technology,Clear Water Bay, Kowloon, Hong Kong\\
}
\date{Accepted XXX. Received YYY; in original form ZZZ}

\pubyear{2020}
\begin{document}
\label{firstpage}
\pagerange{\pageref{firstpage}--\pageref{lastpage}}
\maketitle
\begin{abstract}
\textbf{\textit{Context.}} 
The rate of detection of Fast Radio Bursts (FRBs) in recent years has increased rapidly and getting samples  of sizes  $\mathcal{O}(10^2)$ to $\mathcal{O}(10^3)$ is likely possible. FRBs exhibit short radio bursts in order of milliseconds at frequencies of about $1$ $\mathrm{GHz}$. 
They are bright and have high dispersion measures which suggest they are of extra galactic origin. Their extragalactic origin allows probing the electron density in the intergalactic medium. One important consequence of this is, FRBs can help us in understanding the epoch of helium  reionization. \\
\textbf{\textit{Aims.}} In this project, we tried to explore the possibility of identifying the epoch of Helium II (HeII) reionization, via the observations of early FRBs in range of $z=3$ to $4$. We constrained the HeII reionization with different number of observed early FRBs and associated redshift measurement errors to them.\\
\textbf{\textit{Methods.}} We build a model of FRB Dispersion Measure following the HeII reionization model, density fluctuation in large scale structure, host galaxy interstellar medium and local environment of FRB contribution. We then fit our model to the ideal inter galactic medium (IGM) dispersion measure model to check the goodness of constraining the \ce{HeII} reionization via FRB measurement statistics. \\
\textbf{\textit{Conclusion.}} We report our findings under two categories, accuracy in detection of \ce{HeII} reionization via FRBs assuming no uncertainty in the redshift measurement and alternatively assuming a varied level of uncertainty in redshift measurement of the FRBs. We show that under the first case, a detection of $N\sim\mathcal{O} (10^2)$ FRBs give an uncertainty of $\sigma (z_{r, fit})\sim0.5$  from the fit model, and  a detection of $N\sim\mathcal{O} (10^3)$ gives an uncertainty of $\sigma (z_{r, fit})\sim0.1$. While assuming a redshift uncertainty of level $5-20\%$, changes the $\sigma (z_{r, fit})\sim0.5$ to $0.6$ in $N\sim 100$ case respectively and $\sigma (z_{r, fit})\sim0.1$ to $0.15$ for $N \sim 1000$ case. 
\end{abstract} 


\section{Introduction}

\label{sec:intro}
Fast radio bursts (\acs{FRB}s) are a new sensation in astronomy. They were first detected in 2007 \citep{lor}.
 They are  radio transients of short duration. Observations show, they have high dispersion measure (\acs{DM}) and high galactic latitude $(|b|\textgreater \ang{40})$ \citep{katz} of incidence, thus confirming that they originated at cosmological distances \citep{jaros}. This enables them to be used as an efficient cosmological probe.   Although origin of FRBs are still not known definitively, but it is understood that they are caused by an unknown high energy phenomena \citep{frb3}. On the radio sky, radio transients vary according to the dynamical time. Till now only a very few FRBs  observed are typically repetitive (thus ruling out any possibility of their origins  from cataclysmic events) \citep{frb0,frb1,frb2} and only one is detected exhibiting periodicity of $16.35\pm0.18$ days \citep{frb4}.    For short bursts, the dynamical time ranges from $[0.1\mathrm{ms}-10\mathrm{ms}]$,  corresponding from a neutron star to a white dwarf respectively \citep{fan}. The FRB signals possess interesting features such as:
 \begin{itemize}
     \item FRB signals have a time delay which is inversely proportional to the square of the frequency i.e.  $\Delta t \propto {\nu}^{-2}$ where $\nu$ is the radiation frequency of the burst  \citep{Wiklind}. 
     \item The dependence on the frequency of the burst’s width, which corresponds to Kolmogorov’s power law. \citep{kol}, according to which the burst’s width is proportional to $\nu^{-4}$. Mathematically the broadened width relation is given in terms of the \ce{DM} as,
 \end{itemize}
\begin{equation}
\label{eq3}
width=8.3 \times 10^{-3} \left( \frac{\rm DM}{\rm pc~cm^{-3}} \right)\left(\frac{\Delta \nu}{\rm MHz} \right) \left( \frac{\nu}{\rm GHz}\right)^{-3} {\rm ms},
\end{equation}
where $\Delta \nu$ is the channel bandwidth \citep{hashimoto2019luminosity}.

   Therefore the higher the frequency, less is the time delay. This dispersion feature corresponds to cold plasma and it is predicted that the radio bursts were propagating through such cold plasma. The delay mostly happens due to scattering of the free electrons along the line of sight. This important information is thus encoded in  the redshift (\acs{z}) information of the FRBs. Therefore the magnitude of the integral of the electron density from the source to the observer along the line of sight of the FRB gives the measure of this dispersion called as the \acs{DM}.  \acs{DM} is a time delay of the signal in comparison with the time the signal traveled in vacuum.  The other particles does not interact as much as electrons, their influence is thus insignificant. The general expression for \acs{DM}, therefore contains only the effect of electrons which is calculated as below \citep{lor,deng}:  
\begin{equation}   
\label{e2}
\mathrm{\acs{DM}} = \int^{z'}_{0} \frac{n_{\ce{e}}}{(1+z)} \, dl.
\end{equation}
where $n_e$ is the electron density.

The contribution of the DM by materials on a part of line of sight only from $z'$ to $z''$ range is
\begin{equation}   
\label{e3}
\mathrm{{DM\rvert_{z'}^{z''}}} = \int^{z''}_{z'} \frac{n_{\ce{e}}}{(1+z)} \, dl.
\end{equation}

By calculating \acs{DM} versus redshift, we can use \acs{FRB}s as precision probes of the Universe  \citep{li} especially for studying problems like the missing baryons  \citep{munoz2018finding}, dark energy equation of state  \citep{zhou} and reionization (especially the second helium (\ce{HeII}) reionization)  \citep{linder}. However the available statistics of FRB at our disposal at the moment are too scarce to make elaborate cosmological estimations from them. But  keeping in mind of the future detection scopes \citep{chime}, in this paper we discuss the prospect of using \acs{FRB}s  to investigate their potential in  probing the mechanism of  HeII reionization, In particular, by considering the role of anisotropy of electron distribution in host galaxy  \citep{linder}.

This paper is outlined as follows : in section \S\ref{sec:Reionization} we discuss about the epoch of reionization from the point of view of \acs{FRB} study and this paper. In the next section \S\ref{sec : DM_{frb}}, we summarize the contributions to the \acs{FRB} \acs{DM}  from different  factors. After that the remainder of the paper is focused on  trying to infer  the \ce{FRB} statistics required for  constraining the  $\ce{HeII}$ reionization detection redshift, while taking into consideration realistic redshift uncertainty measurements,  section (\S\ref{sec : fit}). In the last two sections (\S\ref{sec:discus}, \S\ref{sec : Conc}), we summarize our results and provide the final inference based on our analysis. 


\section{\ce{Helium} II  Reionization}
\label{sec:Reionization}

The epoch of reionization in the history of the Universe, was when the first electrons (\ce{e-}) of the neutral hydrogen ($\ce{HI}$) and helium ($\ce{HeI}$) were lost \citep{reioni,reioni2,liddle,peacock} from their outer shells. This epoch marked an important phase in the structure of the Universe rendering the intergalactic medium ionized from neutral. Substantial scientific energy has been put into understanding this process and what triggered it's epoch and it's subsequent evolution. Current constraints strongly suggests that this period occurred within a redshift range of $6< z \sim15$ \citep{barkana,bromm2004first,Paoletti2020}. Post the epoch of reionization, much later in the timeline of the Universe (around two billion years since the Big Bang),  followed the second ionization of the  helium ions,   $\ce{HeI\bond{->}HeII}$. This transition is expected to occur around $z\sim3$ \citep{mcquinn2009he,he2a,he2b,he2c}. This phenomenon is referred to, as the Helium reionization ($\ce{HeII}$).  However decisive observational detections are missing for the $\ce{HeII}$ transition signatures. The strongest detection comes from the far ultraviolet spectra of the $\ce{HeII}$ $\ce{Ly}\alpha$ forest from the lines of sight to several quasars along $z\sim3$ \citep{mcquinn2009he,caleb2019constraining}. However, the comparatively less number of lines of sight, puts a high statistical uncertainty in the measurement of the exact time and nature of this process. There are searches to try to find other methods for identifying this transition with better precision. One of them being  the study of the evolution of the temperature of the intergalactic medium around the neighbourhood of $z\sim3$ \citep{caleb2019constraining,mcquinn2009he}. 

\acs{FRB}'s in this context, could be useful for studying the epoch of the $\ce{HeII}$ reionization via their DM \citep{caleb2019constraining}. This is because \acs{FRB}'s last for short instant (few milliseconds) and this enables the study of all the ionized baryons (integrated column density) along the observed line of sight in it's path.

\section{\acs{DM} of FRB}
\label{sec : DM_{frb}}
The expression of \acs{DM}  from a \acs{FRB} is given in equation \eqref{e2}. 
    The total \acs{DM} from a FRB however, consists of four main contributing factors, namely the \acs{DM} related to the Milky Way $\mathrm{\acs{DM}_{MW}}$, to the Intergalactic Medium $\mathrm{{DM}_{IGM}}$, to the host galaxy $\mathrm{{DM}_{host}}$ and to the source itself   $\mathrm{{DM}_{source}}$ \citep{thornton}, 

\begin{equation}
\label{e1}
\mathrm{\acs{DM}} =\mathrm{DM_{MW}}+ \mathrm{DM_{IGM}+ \mathrm{DM_{host}+\mathrm{DM_{source}}}}
\end{equation}

However, we do not have the details of the FRB source so we have only a set of possible range for the $\mathrm{DM_{source}}$ \citep{deng}. It has been extensively discussed that $\mathrm{DM_{source}}$ and $\mathrm{DM_{MW}}$ both are ignorable \citep{thornton,schnitz} in the overall $\mathrm{DM}$ budget of the FRB. In our case, the $\mathrm{DM_{host}}$ factor is further broken up into two constituent terms, $\mathrm{DM_{host}=DM_{local}+DM_{Galaxy Disk}}$. Where $\mathrm{DM_{local}}$ is the region near the FRB in question with high star formation rate. While the term $\mathrm{DM_{Galaxy Disk}}$ is the term contributing from the galactic disk region with the interstellar medium components. 
\subsection{\acs{DM} Contribution by IGM}
\label{sec: DM_IGM}
\subsubsection{HeII reionization model}
\label{sec: DM_IGM_model}

\citep{caleb2019constraining} derived a general expression for the \acs{DM} estimate from an ionized intergalactic medium. Assuming a universe of purely Helium and Hydrogen, with a Helium mass fraction of Y, the number density of free electron density is given by the following expression : 
\begin{equation}
n_{\rm e}=\frac{\rho_{\rm c, 0}\Omega_{\rm b}f_{\rm IGM}}{m_{p}}\,\left[(1-Y)  \chi_{\rm e,H}(z)+\frac{Y}{4}  \chi_{\rm e,He}(z)\right](1+z)^3
\end{equation}
where $\rho_{\rm c, 0}$ is the critical mass density at $z=0$, $f_{IGM}$ is the fraction of the baryon mass in the IGM (to a first order approximation \citep{fukugita,shull2012baryon} showed this can be approximated to $f_{IGM}=0.83$). $m_p$  refers to the mass of proton. Y refers to the Helium mass fraction, which is measured to be 0.243 by Planck. $\Omega_{\rm b}$ is the current baryon mass fraction of the Universe. $\chi_{\rm e,H}(z),\chi_{\rm e,He}(z)$ are the ionization functions for each species of hydrogen and helium as a function of the redshift $z$.  Combining this with  $dl$ : 
\begin{equation}
dl=\frac{1}{1+z}\frac{c}{H_0}\frac{dz}{\sqrt{\Omega_m(1+z)^3+\Omega_\Lambda}}    
\end{equation}
we get the expression for $\mathrm{DM_{IGM}}$ from eq.(\ref{e2}) as :
\begin{equation}
\mathrm{DM_{IGM}} (z) = \frac{3cH_{0}\Omega_\mathrm{b}f_\mathrm{IGM}}{8\pi G m_{p}} \, \times \, \int_{0}^{z} \frac{f_{e}(z) \, (1+z) \, dz}{\sqrt{\Omega_{m}(1+z)^3 + \Omega_{\Lambda}}}    
\label{e3}
\end{equation}
for a flat Universe, where 
\begin{multline}
f_{e}(z) = (1-Y) \chi_{e,\mathrm{H}}(z) + \frac{Y}{4}  \chi_{\rm e,He}(z)\\
= (1-Y) \chi_{e,\mathrm{H}}(z) + \frac{Y}{4}[\chi_{e,\mathrm{He\textsc{ii}}}(z) + 2\chi_{e,\mathrm{He\textsc{iii}}}(z)]
\label{e4}
\end{multline}
Here $\chi_{e,\mathrm{He\textsc{ii}}}(z)$ and $\chi_{e,\mathrm{He\textsc{iii}}}(z)$ corresponds to the ionization functions of singly ionized Helium HeII and doubly ionized Helium HeIII respectively. We assume a sudden reionization occurring at \acs{$z_r$}, and we express the ionization fraction of Helium as following equations:  
\begin{equation}
\chi_{e,\mathrm{He\textsc{ii}}}(z) =
\begin{cases}
1, & if z>z_r \text{ (before HeII reionization)} \\
0, & if z<=z_r \text{ (after HeII reionization)}\\
\end{cases}
\label{e4}
\end{equation}

\begin{equation}
\chi_{e,\mathrm{He\textsc{iii}}}(z) =
\begin{cases}
0, & if z>z_r \text{ (before HeII reionization)} \\
1, & if z<=z_r \text{ (after HeII reionization)}\\
\end{cases}
\label{e4}
\end{equation}

Before HeII reionization, all helium are singly ionized HeII, so $\chi_{e,\mathrm{He\textsc{ii}}} = 1$ and $\chi_{e,\mathrm{He\textsc{iii}}} = 0$. After HeII reionization, all helium are doubly ionized HeIII, so $\chi_{e,\mathrm{He\textsc{iii}}} = 1$ and $\chi_{e,\mathrm{He\textsc{ii}}} = 0$.

We assume all hydrogen atoms are ionized in our simulation scope, i.e. $\chi_{e,\mathrm{H}}(z) = 1$.
\begin{figure}
\includegraphics[width=\columnwidth]{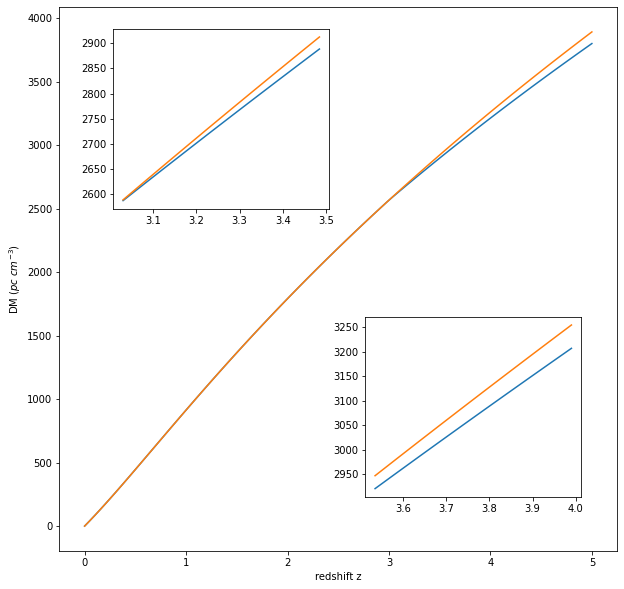}
\caption{Plot of dispersion measure  from $\mathrm{IGM}$ $(\mathrm{DM_{IGM}})$ versus redshift. Blue line indicates \ce{HeII} reionization happened at $z=3$, red line indicates \ce{HeII} reionization happened earlier than scope of plot, i.e. $z>5$. Inset plots show same curve zoomed in at $z=3$ to $3.5$ and $3.5$ to $4$.} 
\end{figure} 




\subsubsection{Fluctuation of $\mathrm{DM_{IGM}}$}
In the above model, IGM is assumed to be even, ionized gas. In reality, large scale structures like galaxy filaments and halos exists, and bring in uncertainty in estimation of $\mathrm{DM_{IGM}}$. From numerical simulation results, $\mathrm{DM_{IGM}}$ is obtained to be  $\sim280$   pc~cm$^{-3}$ at z=1.5 \citep{zheng}\citep{mcquinn2014}\citep{fauch}.

To propagate the uncertainty in $\mathrm{DM_{IGM}}$ to earlier Universe, we consider the variance of baryonic matter at different $z$. After ionization, we assume roughly all atoms in the hot intergalactic medium is ionized, so the electron density should be linearly related to the baryonic density $n_{bar}$. Therefore, we propagate the variance of $\mathrm{DM_{IGM}}$ at some $z$ using a large scale structure simulation: the  \href{https://wwwmpa.mpa-garching.mpg.de/galform/virgo/millennium/}{Millennium} simulation project. 

Statistically, $\sigma^2(\sum_i x_i) = \sum_i \sigma^2(x_i)$, if $x_i$ are uncorrelated. For $\mathrm{DM_{IGM}} = \int^{z}_{0} n_{\ce{e}}/(1+z)dl$, the \acs{DM} comes from integrating the electron density through the path of \acs{FRB} signal travelled. Considering the FRB signal from $z>3$, the distance from source to observer (Earth) is around 2 orders larger than the known large scale structures($\sim100$ $\mathrm{Mpc}$). Since the Universe is assumed homogeneous beyond scale of these structures, we can assume $n_{bar}$ on the line of sight from the FRB to us is roughly uncorrelated.

Now we can write the variance of $\mathrm{DM_{IGM}}$:
$$\sigma^2(\mathrm{DM_{IGM(z)}}) \propto \int_0^z \frac{\sigma^2(n_{bar})}{1+z}dl $$ 
$$\propto \int_0^z\frac{\sigma^2(n_{bar})}{(1+z)^2\sqrt{\Omega_m(1+z)^3+\Omega_\Lambda}}dz$$

From Millennium database:  \href{http://galaxy-catalogue.dur.ac.uk:8080/Millennium/Help?page=databases/millimil/database}{millimil} simulation, we obtained the relative variance of baryonic matter from fitting as a function of $z$:
\begin{equation}
\sigma^2(n_{bar}) = 3.079e^{-1.429z} + 0.6597e^{-0.3328z} 
\end{equation}
Integrating gives

$\frac{\sigma^2\left(\mathrm{DM_{IGM(z=3)}}\right)}{\sigma^2\left(\mathrm{DM_{IGM(z=1.5)}}\right)} = 1.021$ and

$\frac{\sigma^2\left(\mathrm{DM_{IGM(z=6)}}\right)}{\sigma^2\left(\mathrm{DM_{IGM(z=1.5)}}\right)} = 1.025$. 

This tell us that the $\sigma(DM_{IGM})$ is roughly constant at $280$ pc~cm$^{-3}$ after $z=1.5$.

\subsection{\acs{DM} Contribution by host galaxy disk}
\label{sec: DM_galaxy}
From the observed \acs{FRB} \acs{DM} which is greater than the foreground DM, it is understood that their origin is extragalactic. Thus the estimation of the contribution by the host galaxy in the overall \acs{DM} budget  is necessary.  \citep{hostg1} have shown that depending on the type of the galaxy as hosts, the peak \acs{DM} contribution could vary between few thousands (edge-on spiral galaxy) of  pc~cm$^{-3}$ to few tens (dwarf and elliptical galaxies) of pc~cm$^{-3}$. Additionally, based on the the inclination angle of the host galaxy the line of sight of the incident \acs{FRB} will vary and hence  plays an important role in the host galaxy \acs{DM} contribution. 
\subsubsection{Galaxy types}
Host galaxies to FRB's could be modelled based on either, spiral, elliptical or dwarf galaxies. However knowledge of dwarf and elliptical galaxies in terms of their electron densities are not well modelled. Also, presence of local high density clump like regions within a galaxy can enhance the \acs{DM} contribution from the host galaxies, should the line of sight propagate through such regions. Such clumps could be linked to  $\ce{H}\ce{II}$  regions. Such $\ce{HII}$  regions are scarcely observed in elliptical galaxies in comparison to the arms of the spiral galaxies \citep{zhou,hostg2}.   In this analysis, we have used spiral galaxies as a result to model for the host galaxy \acs{DM} contribution. For \acs{DM} contribution from the elliptical or dwarf  galaxies one can consult \citep{hostg1}. Based on simulations  with $10,000$ FRB's they showed that for spiral galaxies the peak \acs{DM} contribution can be of $\mathcal{O}(10^3)$ at high inclination angle ($\textgreater\ang{70}$), while for elliptical and dwarf galaxies on average the peak  \acs{DM} contribution is $37$  pc~cm$^{-3}$ and $45$ pc~cm$^{-3}$ as a function of the inclination angle between $[0, $ $\pi/2]$ respectively.

\subsubsection{Galaxy database as reference}
For our analysis, in order to model the host galaxy \acs{DM} we used the data from the \href{http://astroweb.cwru.edu/SPARC/}{SPARC} database \citep{sparc}. From this database, we used the maximum disk data compiled from $175$ galaxies by \citep{sparc2}. The scale length (R disk) measured at $3.6$ micron band by the Spitzer and the baryonic mass are specifically used.
\subsubsection{\acs{DM} model of disk galaxies}
To model the host galaxy DM, using a spiral galaxy model, we used the distribution function presented by \citep{hostg1}. According to them the behaviour of the \acs{DM} follows a skewed Gaussian distribution, given by :
\begin{equation}
\frac{dN}{d\mathrm{DM}}=N_0~e^{-\frac{(\mathrm{DM}-\xi)^2}{2\omega^2}}~\int_{-\infty}^{\alpha(\frac{\mathrm{DM}-\xi}{\omega})}e^{-\frac{t^2}{2}}dt,
\label{sgdis}
\end{equation}
the parameters in the equations bear the usual representations as mentioned in the original paper. We used this distribution function (equation \ref{sgdis})  together with the SPARC data mentioned above. For computing the parameters, $\xi,\omega,\alpha$ we further made use of the values mentioned in the table $1$ of \citep{hostg1}. We obtained the following  fit values,
\begin{equation}
\xi = 66.52\times\left[e^{\left(-((i-90)/21.02)^2\right)}\right] + 27.56\times\left[e^{\left(-((i-90)/114.4)^2\right)} \right]     
\end{equation}
\begin{equation}
 \omega = 35.36\times\left[e^{\left(0.009625i\right)}\right] + 0.004973\times\left[e^{\left(0.1266i\right)} \right]  
\end{equation}
\begin{equation}
    \alpha = 3.003\times\left[e^{\left(-0.0008232i\right)}\right] + 1.54e-15\times\left[e^{\left(0.3867i\right)} \right] 
\end{equation}
the viewing angle, $i$, is then randomly generated from a flat distribution of $0$ to $90$ degree. 
\subsubsection{Correction factors}
\label{sec: DM_galaxy_corr}
In the above model, a Milky Way like galaxy (they adopted size and mass of Milky Way, with small scale structure neglected) is used to simulate the \acs{DM} of interstellar medium within galactic disk. Milky Way own  \SI{3.6}{\micro\metre}  scale length of $~3.6$ kpc and a baryonic mass of $\sim 1.2\times 10^{11} \ M_{sun}$. To make the model for various disk galaxies, we consider a correction factor as follow.

Assume mean electron density is directly correlated with baryonic density within a galaxy, and all disk galaxies share same shape as milkyway as a simplified model,  $\langle n_e \rangle \propto m/r^3$, 
$n_e$ should have a correction factor of 
\begin{equation}
\frac{m}{m_{MW}}\left(\frac{r_{MW}}{r}\right)^3 = \frac{m}{1.2\times 10^{11}}\left(\frac{3.6}{r}\right)^3.
\end{equation}
From eq.\eqref{e1} $\mathrm{DM} = \int^{z}_{0} n_{\ce{e}}/(1+z) \, dl.$ , We should consider an additional factor of $r$ for DM since the path length of \acs{FRB} signal travelling inside the galaxy disk is also affected by size of galaxy. In total we need to multiply the \acs{DM} from above model by a factor of 
\begin{equation}
\frac{m}{m_{MW}}\left(\frac{r_{MW}}{r}\right)^3\left(\frac{r}{r_{MW}}\right) =\frac{m}{1.2\times 10^{11}}\left(\frac{3.6}{r}\right)^2.
\end{equation}

\subsection{\acs{DM} Contribution by local environment of \acs{FRB}}
\label{sec: DM_local}
\subsubsection{Giant star forming regions}
Recently, a non-repeating \acs{FRB}, 180916.J0158+65, is localized to a star forming region inside spiral arm of a nearby spiral galaxy \citep{marcote2020repeating}. \acs{FRB} 181112 is also located to a active star forming galaxy \citep{prochaska2019low}. A research on 21 \acs{FRB}s show that the host galaxies contribute a large mean \acs{DM} of $\sim270$  pc~cm$^{-1}$ \citep{yang2017large}, which possibly comes from nearby plasma like star forming \ce{HII} regions. It is reasonable to assume a significant portion of \acs{FRB}s to correlate with active star forming regions, especially the giant \ce{HII} regions
\subsubsection{HII region model in spiral galaxy}

To estimate possible contribution of \acs{DM} by host galaxy, we construct a simplified model in  which the \acs{FRB} is embedded in a \ce{HII} region, and $\mathrm{DM_{host}}$ is completely contributed by the free electrons within the \ce{HII} region. Since the shape and electron density distribution of the \ce{HII} region is unknown, we assume a spherical, homogeneous \ce{HII} region for estimation. Recalling eq.\eqref{e3}, since the size of \ce{HII} region is negligible in cosmological scale, we can assume the redshift, z, as constant and rewrite the equation as 
\begin{equation}
\mathrm{DM_{host}} = \frac{1}{1+z}\times \left< n_e \right> \times P_L
\label{e5}
\end{equation}
Where $\left< n_e \right>$ is the mean electron density in the \ce{HII} region and $P_L$ is the path length of \acs{FRB} pulse travelled inside the \ce{HII} region. 

The electron density $n_e$ can vary in different \ce{HII} region. To estimate this term, we reference to a research on size and electron density of \ce{HII} regions in nearby galaxy M51 \citep{gutierrez2010galaxy}. We  found a general form of \ce{HII} region mean electron density from fitting the data in M51 model:
\begin{equation}
\left< n_{e,M51} \right> = \begin{cases} 
45.8e^{-r/h}R^{-0.55}cm^{-3},& r<1.4 \mathrm{kpc} \text{ or }4.6 \mathrm{kpc}\\
28.9R^{-0.55}cm^{-3},&r\in [1.4, 4.6] \mathrm{kpc}
\end{cases}
\label{e6}
\end{equation}
Here $h$ represents the scale length of the galaxy in kpc from fitting, $r$ represents the distance of \ce{HII} region from the center of the galaxy in kpc, and $R$ represent the equivalent radius of the \ce{HII} region in pc. In the paper, $h\sim 10$ kpc for M51, which match its neutral hydrogen scale length \citep{gutierrez2010galaxy}. 

In this paper, we modelled our  \acs{FRB}'s host  galaxies based on the M51 model and therefore we use the same model for fitting. In the galaxy's database, we get the scale length of various galaxies at infrared ($3.6\mu m$) band, which  mainly represents the stellar mass instead of neutral hydrogen. The M51 $3.6\mu m$ scale length is measured in \citep{leroy} as 2.8kpc instead of 10kpc of its neutral hydrogen scale length. To account for this difference, we multiply the scale length by a factor of

\begin{equation}
\frac{\text{M51 } 3.6\mathrm{\mu m}\text{ scale length}}{ \text{M51 neutral hydrogen scale length }} = 0.28
\end{equation}

To calibrate the effect of varying density in each galaxy, the electron density is also multiplied by a density factor of $m/r^3$ like in section \S \ref{sec: DM_galaxy_corr}. Here we take the baryonic mass of M51 (HI, \ce{HII} and stellar mass)) as $5\times10^{10}M_{sun}$ from \citep{walter2008things}, \citep{leroy2008star} and \citep{hughes2013probability}.
The overall \ce{HII} region electron density is modelled as follow:
\begin{equation}
\left< n_{e} \right> = \begin{cases} 
45.8e^{-(r/0.28r_{char})}R^{-0.55}\frac{m}{5\times10^{10}}\left(\frac{2.8}{r_{char}}\right)^3,\qquad&\\
\hfill if \frac{2.8r}{r_{disk}}<1.4kpc \text{ or }>4.6kpc\\
28.9R^{-0.55}\frac{m}{5\times10^{10}}\left(\frac{2.8}{r_{char}}\right)^3,\qquad&\\
\hfill if  \frac{2.8r}{r_{char}}\in [1.4, 4.6]kpc
\end{cases}
\label{e6}
\end{equation}
Here \acs{$r_{disk}$} refers to the scale length of the spiral galaxy at $3.6$ $\mu m$ in kpc, $m$ refers to mass of galaxy in $M_{sol}$. As above, $r$ represents the distance of \ce{HII} region from center of galaxy in kpc, and $R$ represent the equivalent radius of the \ce{HII} region in pc. 

From the above assumption, we can randomly generate the $\mathrm{DM_{host}}$ distribution of \acs{FRB}s with known redshift z. 
Observing data in \citep{alvarez2015comprehensive}, \citep{mayya1994embedded} and \citep{arsenault1989circumnuclear}, radius $R$ of giant \ce{HII} regions in nearby galaxies has a $\log$-normal like distribution,
\begin{equation}
R\sim10^{N(\mu = 2.1, \sigma^2 = 0.09)}
\end{equation}
where $N$ represents the usual  normal distribution. For generation of $r$ i.e. the  position of \ce{HII} regions from the center of the galaxy, we use an absolute normal distribution of $\left|N(\mu = 0, \sigma^2 = r_{disk}^2)\right|$, motivated by the fact that  \acs{$r_{disk}$} is the scale length of the galaxy in $3.6$ $\mu m$ which represents mainly stars and dust.

After this, we need to know the  path length $l$ of the \acs{FRB} signal travelled inside the \ce{HII} region. 

Assuming that the \acs{FRB} is generated in an uniform random position inside a spherical \ce{HII} region, the path length of \acs{FRB} signal travelled inside the \ce{HII} region $l$, is calculated by
\begin{equation}
P_L = z_c + \sqrt{R^2 - x_c^2 - y_c^2} 
\end{equation}
Where $(x_c,y_c,z_c)$ is the  coordinate of \acs{FRB} with respect to the center of \ce{HII} region, in pc, and R is the radius of the \ce{HII} region. The equation follows a simple geometric model, shown in figure \ref{model00}:
\begin{figure}
\includegraphics[width=1.1\columnwidth]{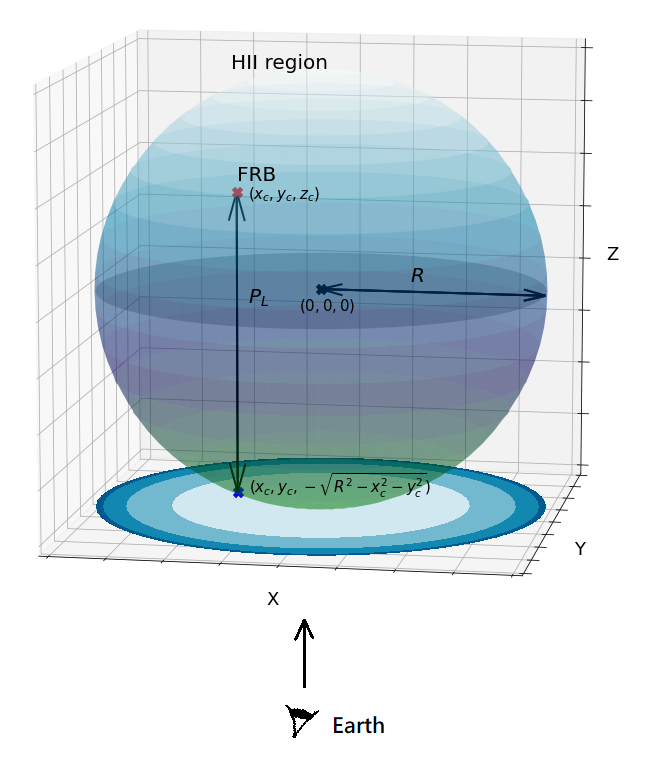}
\caption{Model of FRB embedded in \ce{HII} region and the projected path length of the signal travelled inside the \ce{HII} region, $P_L$. Red cross represents FRB, blue cross represents the corresponding point on surface of spherical \ce{HII} region. Earth (observer) is along the  $+z$ direction} 
\label{model00}
\end{figure}

By running the above model from section 3.2 and 3.3, we obtained a heavy tailed contribution of DM host galaxy and FRB local environment with mean of $\sim 275$ pc $cm^{-3}$, which agreed with the observed value of $270$ pc $cm^{-3}$ from 21 known FRBs \citep{yang2017large}. The corresponding variance can reach upto  $\sim 390$ pc $cm^{-3}$ from these terms. 

For a \ce{FRB} at redshift z, the expected contribution of $\mathrm{DM_{HII}}+\mathrm{DM_{GalaxyDisk}} = 275$ pc $cm^{-3}/(1+z)$.
\subsection{\acs{DM} Contribution by Milky Way and nearby Universe} 

\begin{figure*}[h!]
\includegraphics[width=\textwidth]{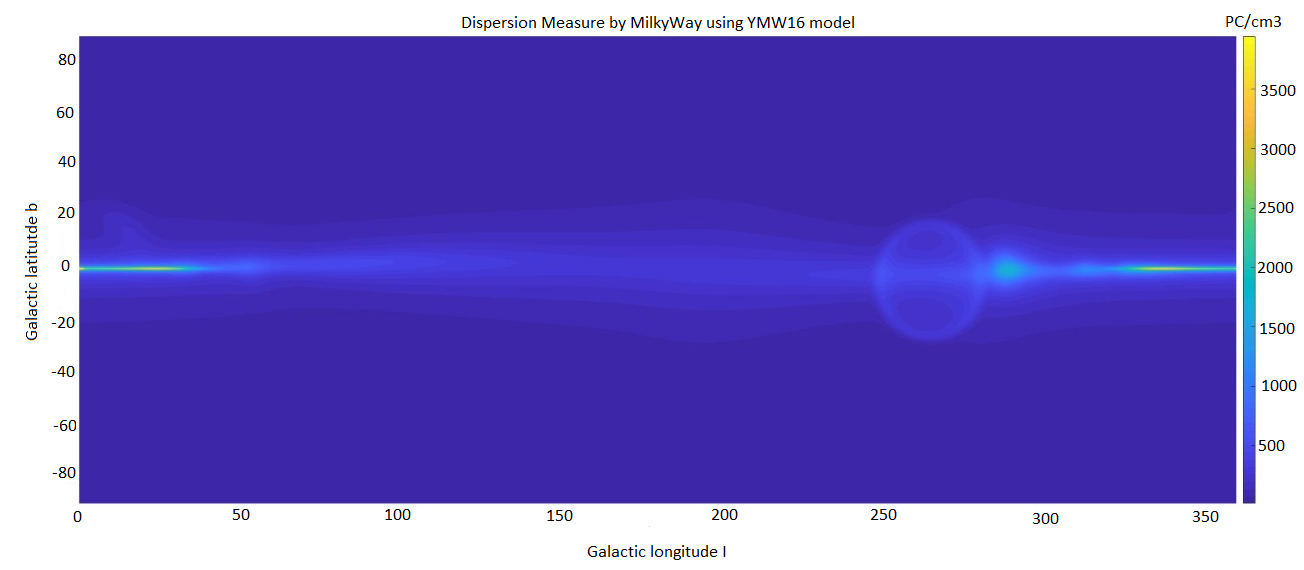}
\caption{Calculated disperion measure (DM) for Milky Way  using $\mathrm{YMW16}$ electron density model} 
\label{C1}
\end{figure*}

We also calculate the foreground contribution from the Milky Way’s disk and spiral arms using the widely-used  \href{http://119.78.162.254/dmodel/index.php}{$\mathrm{YMW16}$} distribution \citep{yao}. $\mathrm{YMW16}$ is a model for the distribution of free electrons in the Milky Way, Magellanic Clouds and nearby IGM. The model is constructed based on DM measurement on radio pulsars. This model gives a good description of the Milky Way structure including the spiral arms and central bulge. The model also includes some known \ce{HII} regions in our galaxy, like the Gum Nebula \citep{gum}, Galactic Loop I \citep{loop} and the Local Bubble \citep{bubble}. Figure \ref{C1} shows calculated DM for Milky Way from the earth using $\mathrm{YMW16}$ electron density model. Therefore, we can deduce $\mathrm{DM_{MW}}$ later from our total DM to locate FRB to fair resolution.

\label{sec: DM_MW}

\subsection{Overall \acs{DM} and the Variance}

\begin{figure}
\includegraphics[width=\columnwidth]{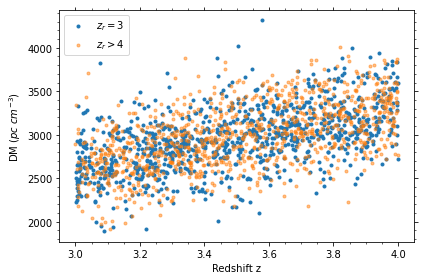}
\caption{ Dispersion Measure (\ce{DM}) distribution as a function of the redshift, of the synthetic \ce{FRB}s generated from the above model, in range of $z=3$ to $4$. The legends show the corresponding reionization redshift $(z_r)$ assumed.  } 
\label{dm_fit}
\end{figure}

From the above model in section \ref{sec: DM_IGM}, \ref{sec: DM_galaxy} and \ref{sec: DM_local}. The overall \acs{DM} of FRB are simulated as follow graph, with 1000 fictitious FRBs plotted (figure \ref{dm_fit}). 

\label{sec: DM_SUM}

\section{Fitting occurrence of HeII reionization with simulated \acs{FRB} \acs{DM} and its variance} \label{sec : fit}
With the simulated \acs{FRB}s versus redshift (z) distribution, along with the availability of the models of the individual \ce{DM} components, we can now statistically analyse the number of FRB in range of $z=3$ to $4$ needed to constraint the epoch of \ce{HeII} reionization. 

We generate sets of synthetic FRBs with DM uncertainty generated from the above model, then fit them back to the ideal $\mathrm{DM_{IGM}}$ model as described in section \ref{sec: DM_IGM_model}. $\langle\mathrm{DM_{host}}\rangle$ of $275$  pc $cm^{-3}$ is deduced from data to calibrate the effect of $\mathrm{DM_{host}}$. The synthetic \ce{FRB}s are grouped into bins of various sizes, so we can analysis the effect of the accuracy of fitting as a function of the number of \ce{FRB}s observed. Additionally we also consider the error in the measured redshift of each FRB. We introduced a Gaussian uncertainty in redshift measurement $z_{FRB}$ from $0\%$ to $20\%$ level, i.e. $\sigma(z_{FRB}) = G\times z_{FRB}$, where $G$ range from $0\%$ to $20\%$. 

Standard deviation of  \acs{$z_{r, fit}$}, $\sigma (z_{r, fit})$ is obtained from the fitting, where  $z_{r,fit}$ is the redshift of \ce{HeII} reionization epoch obtained from the fit and the corresponding $\sigma (z_{r, fit})$ is the associated uncertainty. A smaller $\sigma (z_{r, fit})$ means that we have higher confidence on \ce{HeII} reionization happening at \acs{$z_{r, fit}$}. Taking a ground truth value of HeII reionization happening at $z=3$, the  uncertainty $\sigma (z_{r, fit})=\epsilon$ then tells us that there is a $1 - \sigma$ confidence from the fit that \ce{HeII} reionization happened in the neighbourhood of  $z=3.0\pm\epsilon$. 

\section{Discussion}
\label{sec:discus}
\begin{figure}
    \centering
    \includegraphics[width=\columnwidth]{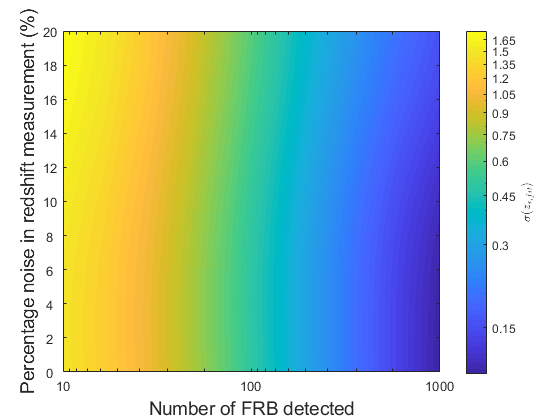}
    \captionof{figure}{Joint plot showing  the uncertainty in the \ce{HeII} reionization detection  as a function of both the \ce{FRB} detection statistics and the associated $\%$ of noise in the corresponding redshift measurement. }
    \label{fig:color_plot_main}
\end{figure}

From the figure \ref{fig:color_plot_main}, we can see the $1-\sigma$ constraint, $\sigma (z_{r, fit})$, tightens when number of \ce{FRB}s increase (left to right variation along the x-axis). While the effect of increasing the noise in measured redshift of each FRB (vertical trend) is sub dominant. 

\begin{figure}
    \centering
    \includegraphics[width=\columnwidth]{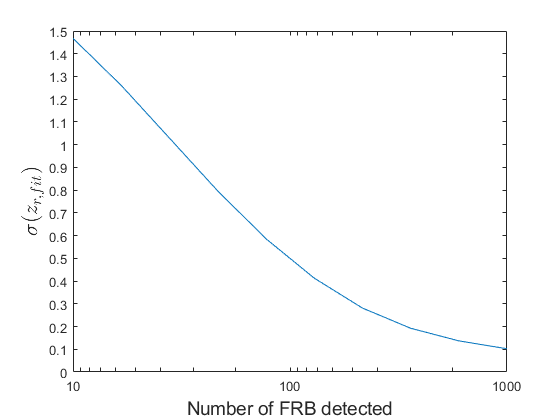}
    \captionof{figure}{Figure showing the uncertainty level $\sigma (z_{r, fit})$ in the \ce{HeII} reionization detection as a function of the \ce{FRB} detection statistics. The above case is presented, while assuming there is no uncertainty involved in the redshift measurement. }
    \label{fig:no_noise}
\end{figure} 

In figure \ref{fig:no_noise}, we plot the \ce{HeII} reionization detection uncertainty as a function of the number of \ce{FRB}s detected, while assuming no error in redshift measurement. It is seen that we can constrain the \ce{HeII} reionization to an uncertainty level of $\sigma (z_{r, fit})\textless 0.5$ for $N_{FRB}\ge 100$ detected. Therefore, to obtain a fair constrain on the \ce{HeII} reionization,  we need hundreds of FRBs detected in $z=3$ to $4$.

\begin{table}
\centering
 \begin{tabular}{||c |c | c||} 
 \hline
 Number of FRB detected & $\sigma(z_{FRB})$ & $\sigma (z_{r, fit})$ \\ [0.5ex] 
 \hline\hline
& 0 &1.47   \\ 
10  & $10\%$ &1.54   \\ 
& $20\%$&1.78   \\ 
 \hline
& 0 &0.500   \\ 
100  & $10\%$ &0.515   \\ 
& $20\%$ &0.602   \\ 
 \hline
& 0 &0.102   \\ 
1000  & $10\%$ &0.114   \\ 
& $20\%$ &0.156   \\ 
 \hline
 \end{tabular}
 \caption{Table summarizing the effect on the constraint of \ce{HeII} reionization uncertainty level as a function of the FRB statistics and the redshift uncertainty. We present scenarios with three  different levels of uncertainty fraction in the measured redshift,  $\sigma(z_{FRB})=[0,10\%,20\%]$ of the FRBs for each of the three cases of $[10,100,1000]$ FRB detections.  It is clear, at this level, we require  $\mathcal{O}(10^2)$ FRBs for a detection of \ce{HeII} reionization around the neighbourhood of $z\sim[3-4]$.}
\label{tab:result}
\end{table}



\begin{figure}
    \centering
    \includegraphics[width=\columnwidth]{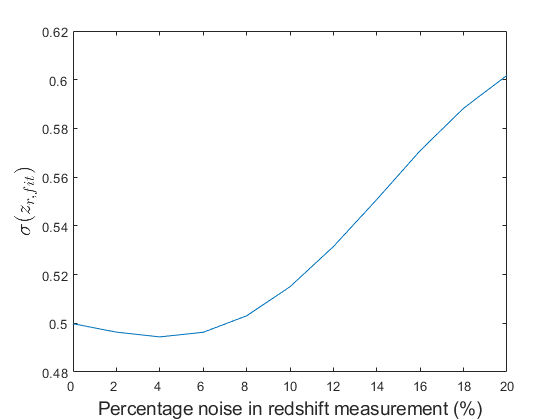}
    \captionof{figure}{Plot showing the effect of redshift uncertainty (varied between $0-20\%$) on the \ce{HeII} reionization detection  for $100$ detected \ce{FRB} case.}
    \label{fig:100FRB}
\end{figure}

 We also show the sole effect on \ce{HeII} reionization detection from the redshift measurement uncertainty by fixing the number of \ce{FRB}s detected at $100$. We can investigate the effect of uncertainty in redshift measurement ($\sigma(z_{FRB})$) in the detection,  as shown in figure \ref{fig:100FRB}. We can see that $\sigma(z_{FRB})$ starts to effect the fitting results when it reaches $\sigma(z_{FRB}) = 6\%\times z_{FRB}$ (x axis, figure \ref{fig:100FRB}). $\sigma (z_{r, fit})$ grow from 0.5 to 0.6 when a percentage noise is turned up from $6\%$ to $20\%$. 

\begin{figure}
    \centering
    \includegraphics[width=\columnwidth]{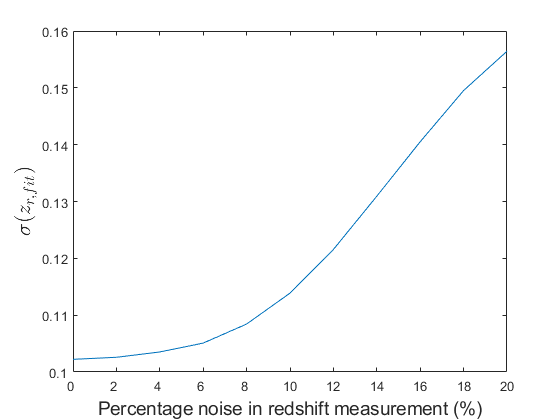}
    \captionof{figure}{Plot showing the effect of redshift uncertainty (varied between $0-20\%$) on the \ce{HeII} reionization detection for $1000$ detected \ce{FRB} case.}
    \label{fig:1000FRB}
\end{figure}

Similar effect is observed when we push the detected FRBs to $1000$, as shown in figure \ref{fig:1000FRB}. $\sigma (z_{r, fit})$ grow from $0.1$ to $0.16$ when a percentage noise of $20\%$ is introduced to $\sigma(z_{FRB})$. 

\section{Conclusion}
\label{sec : Conc}
In this paper, we presented a cosmological based model on constraining the \ce{HeII} reionization redshift in range of $z=3$ to $4$ by detecting the dispersion measure of distant \ce{FRB}s. The model considers contribution from uneven IGM distribution, a host galaxy of disk type and a FRB local environment of giant \ce{HII} region. Data used includes the the millennium simulation project(Millimil database), electron density model of disk type galaxy was based on Milky Way measurement, \ce{HII} region electron density model from observations,  the simulated dispersion measure from host galaxy and local environment fits of the observed data from 21 \ce{FRB}s. 

Synthetic FRBs are generated with the above model in different batch sizes, assuming HeII reionization happens at $z=3$. To simulate real observations, we also considered redshift uncertainty $\sigma(z_{FRB})$  in the measurement of FRB, from 0 to $20\%$ level of  $z_{FRB}$ following a Gaussian noise. 

These FRBs are fitted to an ideal model of HeII reionization as mentioned in section \S\ref{sec: DM_IGM_model}  with least-square fitting. 

From the fitting result, as shown in table \ref{tab:result}, at least around a hundred FRBs in $z=3$ to $4$ is necessary to constrain the epoch of \ce{HeII} reionization, $\sigma (z_{r, fit})$  $\sim 0.5$. With the addition of $\sigma(z_{FRB}) = 20\%z_{FRB}$, $\sigma (z_{r, fit})$ worsens to $0.6$. For a larger population of $1000$ \ce{FRB}s measured, we can constraint  $\sigma (z_{r, fit})$  $\sim 0.1$ ideally, and addition of $\sigma(z_{FRB}) = 20\%z_{FRB}$ worsens this to $\sim 0.16$. 

At the time of completion of this paper, we received an exciting news that a Milky Way soft gamma ray repeater (SGR), SGR 1935+2154, flared a FRB-like, double millisecond pulses with 30ms interval at 28 April 2020. Together there exist X-ray pulses arriving 8.63s earlier, completely match the delay brought by dispersion measure. This is the first time a FRB (or similar event) related to a known source, and provide us hints on possible origin of FRBs. 

The estimation above is still valid since we assumed FRBs origined from source within galaxies, and correlated to young star / pulsar population. Also, SGR has possible correlation with star forming regions: SGR1806-20 is embedded in Westerhout 31, a star forming complex \citep{corbel1997distance}. We hope to perform more detailed analysis in near future base on knowledge on magnetstars.

\section{Acknowledgement}
AM did this work with the grant from RK MES grant AP05135753, Kazakhstan. 
\label{sec:con}

\bibliographystyle{mnras}
\bibliography{reference} 

\appendix

\end{document}